\documentclass[aps,pra,superscriptaddress,amsfonts,amsmath,amssymb,twocolumn,showpacs,floatfix]{revtex4-1}

\usepackage{graphicx}
\usepackage{amsmath,amssymb}

\usepackage{hyperref}
\hypersetup{
        colorlinks=true,
}

\usepackage[utf8]{inputenc}
\usepackage{xcolor}
\newcommand{\be}{\begin{equation}}
\newcommand{\ee}{\end{equation}}

\begin{document}

\title{Density redistribution effects in fermionic optical lattices}

\author{Medha Soni}
\affiliation{Laboratoire de Physique Th\'eorique UMR-5152, CNRS and Universit\'e de Toulouse, F-31062 France}
\affiliation{Theoretische Physik, ETH Zurich, 8093 Zurich, Switzerland}

\author{Michele Dolfi}
\affiliation{Theoretische Physik, ETH Zurich, 8093 Zurich, Switzerland}

\author{Matthias Troyer}
\affiliation{Theoretische Physik, ETH Zurich, 8093 Zurich, Switzerland}
\affiliation{Quantum Architectures and Computation Group, Microsoft Research, Redmond, WA (USA)}
\affiliation{Microsoft Research Station Q, Santa Barbara, CA (USA)}
\date{\today}

\begin{abstract}
We  simulate a one dimensional fermionic optical lattice to analyse heating due to non-adiabatic lattice loading. Our simulations reveal that, similar to the bosonic case, density redistribution effects are the major cause of  heating in harmonic traps. We suggest protocols to modulate the local density distribution during the process of lattice loading, in order to reduce the excess energy. Our numerical results confirm that linear interpolation of the trapping potential and/or the interaction strength is an efficient method of doing so, bearing practical applications relevant to experiments.
\end{abstract}

% 02.70.-c	Computational techniques; simulations
% 03.65.Ud	Entanglement and quantum nonlocality (e.g. EPR paradox, Bell's inequalities, GHZ states, etc.) (for entanglement production and manipulation, see 03.67.Bg; for entanglement measures, witnesses etc., see 03.67.Mn; for entanglement in Bose-Einstein condensates, see 03.75.Gg)

% 37.10.Jk	Atoms in optical lattices

% 67.85.De	Dynamic properties of condensates; excitations, and superfluid flow
% 67.85.Hj	Bose-Einstein condensates in optical potentials
% 67.85.Lm	Degenerate Fermi gases
% 71.27.+a	Strongly correlated electron systems; heavy fermions

\maketitle

%%%%%%%%%%%%%%%%%%%%%%%%%%%%%%%%%%%%%%%%%%%%%%%%%%%%%%%%%%%%%%%%%%%%%%%%%%%%%%%%%%%%%%%%%%%%%%%%%%%%%%%%%
\section{Introduction}
Ultracold atoms in optical lattices provide a versatile toolbox for the realization of strongly correlated quantum Hamiltonians by virtue of their tuneability and controllability~\cite{Bloch2012}. They allow the probing of observables such as magnetic correlations that are particularly interesting for fermionic systems where magnetic ordering arises due to exchange couplings between different spin components. Despite recent progress that has been able to capture the short range physics~\cite{Greif2013,Hart2015}, observing real long-range magnetic correlations is still a great open challenge, because of the low temperatures required for magnetic ordering~\cite{lewen12}.

Without an optical lattice evaporative cooling easily reaches temperatures lower then $T/T_F \approx 0.08$~\cite{Jordens2008,Schneider1520}, but such low temperatures have not yet been achieved in optical lattices. In principle the process of lattice loading should be performed adiabatically, but in practice one will always do so in a finite time, thus deviating from the completely adiabatic regime and incurring some heating.

The breakdown of the adiabatic lattice loading for optical lattices has been well investigated~\cite{wolf2014,mckay11}. Optimizing ramp shapes~\cite{Jakub09}, fast-forward loading schemes~\cite{Masuda14} where an auxiliary potential assists in lattice loading, modulating trap frequency and shape during loading~\cite{Ma2008,bernier09,Paiva2011,dolfi2015},  starting from a low entropy interacting state~\cite{Prasad14}, introducing compensating laser beams~\cite{Mathy2012}, using disordered potentials~\cite{Paiva2015}, Peltier cooling~\cite{Grenier14} are some ways to overcome effects of non-adiabaticities and achieve lower temperatures in optical lattices.
Non-interacting fermions have been studied within superlattice geometries in the continuum~\cite{Kollath15} for both homogenous and trapped set ups. Density redistribution causes population of higher Bloch bands but can be handled by optimizing the initial part of the loading schedule until the gap to higher Bloch bands opens up. Another way to cool down Fermi gases in a deep optical lattice is to use a Bose-Einstein condensate gas as a reservoir to transfer the excess entropy per particle~\cite{Ho09}.

Numerical studies based on single-band models inherently assume a deep optical lattice. However, loading starts from the regime of shallow (or no) lattice, and the important initial phase is thus not captured by single band models. 
Our approach is based on a continuum model which is valid also when the lattice is turned off, and thus describes the entire lattice loading process.

The study of non-adiabaticities in lattice loading and novel cooling schemes are important to achieve the desired low temperatures in experiments. In Ref.~\cite{dolfi2015} some of the authors studied a system of bosons in a 1D optical lattice in the continuum description. For homogenous systems without a confining trap only minimal heating effects, less than $1\%$ of the effective hopping, were encountered even for reasonably short ramp times. Moreover the heating was seen to decrease significantly as longer ramp times were considered. In contrast, when the Hamiltonian included a harmonic confinement potential, significant heating was observed and seen to be more or less constant with ramp time. Significant differences in the density distribution with and without an optical lattice require major density redistributions, leading to heating due to non-adiabaticity. This issue was overcome by dynamically reshaping the trapping potential  during the process of lattice loading in order to reduce the need for redistribution of particles in the lattice. 

\begin{figure}
\include{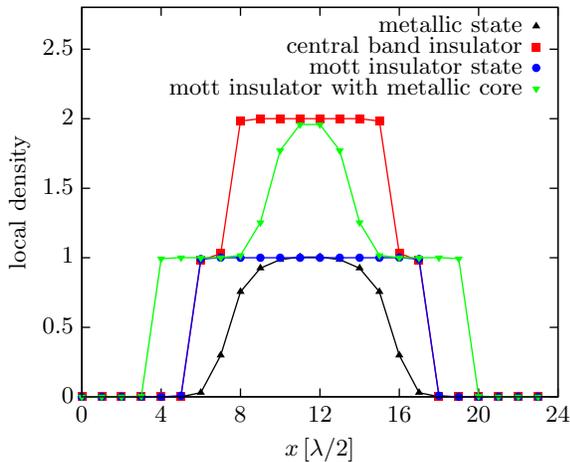}
\caption{Local density distribution of the target states integrated over each unit cell.}
\label{fig:target_states}
\end{figure}

In this paper we generalize this study to spin-1/2 fermions, considering four different \emph{target states}: $(i)$ a metallic state throughout the trap, $(ii)$ a band insulator in the trap center, surrounded by metallic and Mott-insulating regions $(iii)$ a  Mott insulator with unit filling in the trap center $(iv)$ a central metallic core with density larger than one surrounded by a Mott insulating regions. The local density distribution of all these target states are shown in Fig. \ref{fig:target_states}.

We initially show that for the metallic state and the band insulating core, a simple adiabatic ramp  shows negligible defects, whereas the other states, the pure Mott insulator and one with a metallic core, suffer from heating during lattice loading, caused by density defects. We then present revised loading protocols that allow better redistribution of particles during the ramping.
Similar to the bosonic case we dynamically change one or more parameters of the system during loading in order to reduce density defects. We show that this can be achieved in a number of ways, either by dynamically changing the trapping potential as in the bosonic case, or by tuning the interaction during loading.

%%%%%%%%%%%%%%%%%%%%%%%%%%%%%%%%%%%%%%%%%%%%%%%%%%%%%%%%%%%%%%%%%%%%%%%%%%%%%%%%%%%%%%%%%%%%%%%%%%%%%%%%%
\section{Model and Method}\label{sec:model}
We consider a continuum model of spin-1/2 fermions with contact interaction. It can be written in as a function of the field operators $\hat{\psi}^{\dagger}_{\sigma}(x)$ that create a fermions with mass $m$ and spin $\sigma$ at the position $x$. The corresponding annihilation operator is $\hat{\psi}_{\sigma} (x)$. The Hamiltonian of a 1D system of size $L$ can then be written as:

\begin{align}\label{eq:cont_hubbard}
\mathcal{H} &= \sum_{\sigma}\int_0^L  \, dx \, \hat{\psi}^{\dagger}_{\sigma}(x) \bigg[ -\frac{\hbar^2}{2m} \frac{d^2}{dx^2} + V(x)\bigg]\hat{\psi}_{\sigma}(x)& \nonumber \\
&+\frac{g}{2} \sum_{\sigma \sigma'} \int_0^L  \, dx \, \hat{\psi}^{\dagger}_{\sigma}(x)  \hat{\psi}^{\dagger}_{\sigma'}(x)  \hat{\psi}_{\sigma'}(x)  \hat{\psi}_{\sigma}(x),&
\end{align}
where the first term is the kinetic energy and the second one is a site dependent external potential energy. The four-operator term is the contact interaction characterized by the interaction strength $g$, which is obtained from the single particle scattering length~\cite{olshanii1998}.  The external potential carries the potential created by the interfering lattice beams along with the harmonic trap used for confining the system. It is given by 
\be
V(x) = V_0 \cos^2(kx) + \frac{1}{2}m \omega^2 x^2,
\ee
where $V_0$ is the lattice depth, $k = \frac{2\pi}{\lambda}$ is the wave vector of the laser beam and $\omega$ is the frequency of the harmonic trap. The natural energy scale in the problem is the recoil energy defined as $E_r= \frac{\hbar^2 k^2}{2m}$. Our results will be presented in units of $E_r$. 

To simulate the continuum model we discretize space with $N_{\rm discr}=16$ grid points per unit cell with length $a=\lambda/2$ of the optical lattice. This gives a grid spacing  $d = a/N_{\rm discr}$. The continuum Hamiltonian is thus mapped to a Hubbard model written in terms of creation and annihilation operators $c^{\dagger}_{i,\sigma}$ and $c_{i,\sigma}$ respectively, $i$ being the grid site index and $\sigma$ is the spin of the fermion. Its Hamiltonian is
\begin{align}
\mathcal{H} =& -J(d) \sum_{\sigma}\sum_{<ij>} c^{\dagger}_{i,\sigma} c_{j,\sigma} + {\rm h.c.}  \nonumber\\
&+ \sum_{\sigma\sigma'}\sum_{i} \frac{U(d)}{2}c^{\dagger}_{i,\sigma}c^{\dagger}_{i,\sigma'}c_{i,\sigma'}c_{i,\sigma} \nonumber \\
&+ \sum_{\sigma}\sum_{i} \epsilon_i(d) n_{i\sigma},
\end{align}
where the kinetic term becomes the hopping amplitude $J(d)=(\hbar^2 / 2m) / d^2$ between adjacent grid sites $i$ and $j$, the contact interaction turns into an on-site interaction $U(d)=g / d$ and the external potential is implemented as a site-dependent chemical potential $\epsilon_i(d) = V(d/2 + i\,d) + 2 (\hbar^2 / 2m) / d^2$.

Note that in contrast to the effective single-band Hubbard model valid in a deep optical lattice with  $N_{\rm discr}=1$ lattice sites per unit cell, our model corresponds to an effective  $N_{\rm discr}$-band model, which allows the accurate simulation of the initial loading regime with no or very shallow optical lattices.

%%%%%%%%%%%%%%%%%%%%%%%%%%%%%%%%%%%%%%%%%%%%%%%%%%%%%%%%%%%%%%%%%%%%%%%%%%%%%%%%%%%%%%%%%%%%%%%%%%%%%%%%%
We simulate a fermionic optical lattice model numerically with the density matrix renormalization group method (DMRG)~\cite{white1992,schollwock2011}.  DMRG is based on a variational ansatz wave function called matrix product state (MPS), which for one-dimensional quantum systems reduces the exponentially growing complexity to just a polynomial scaling by limiting the amount of entanglement which is captured by the ansatz. The accuracy of the algorithm is systematically improved with an increase of the MPS bond dimension $M$.

 To overcome convergence problems of the standard DMRG approach in large dilute lattices we use the multigrid DMRG algorithm~\cite{dolfi2012}. Time evolution within the MPS framework is performed making use of the time-dependent variants of DMRG \cite{vidal2003,daley2004,white2004}, which split non-commuting terms in the unitary time evolution operator via a second-oder Suzuki-Trotter decomposition on a small time step $\delta_t=0.01\, \hbar / E_r$. As our goal is to evolve the system being as adiabatically as possible, a modest bond dimension between $M=400$ and $M=600$ turned out to be sufficient.

Note that due to the presence of a trapping potential the open boundary conditions of standard DMRG simulations do not introduce any errors, as long as we keep the system size $L$ larger than the effective size $L_{\rm eff}$ of the trapped fermionic cloud.

%%%%%%%%%%%%%%%%%%%%%%%%%%%%%%%%%%%%%%%%%%%%%%%%%%%%%%%%%%%%%%%%%%%%%%%%%%%%%%%%%%%%%%%%%%%%%%%%%%%%%%%%%
\section{Results}\label{sec:results}

\subsection{Lattice loading protocols and observables}\label{sec:results_summary}
To simulate optical lattice loading we first calculate the ground state wave function $|\psi_{\rm init}\rangle$ in the absence of an optical lattice \textit{i.e.} $V_0(t=0)=V_i=0$. 
This state $|\psi(t)\rangle$ is then evolved under a time-dependent Hamiltonian with lattice potential $V_0(t)$. In our simulations we use a linear ramp that interpolates between the initial depth $V_i$ and  final depth $V_f$ of the optical lattice as:
\be
V_0(t) = V_i + (V_f - V_i) \, \frac{t}{t_R},
\ee
where  $t_R$ is total ramp time. At the end of the lattice loading the model is expected to have reached the target state with lattice potential $V_0(t_R)=V_f =8\, E_r$. 
The final state $|\psi_{\rm final}\rangle \equiv |\psi(t = t_R)\rangle$ is  then compared to the target ground state $|\psi_{\rm target}\rangle$.

To quantify and understand the origin of the defects we  calculate several observables during the evolution of the wave function $|\psi(t)\rangle$ to the final state $|\psi_{\rm final}\rangle$. 
Of particular interest are the excess energy per particle 
\begin{equation}
q\bigr|_{(t = t_R)} = {\big(E [|\psi_{\rm final}\rangle] - E[|\psi_{\rm target}\rangle]\big)}/{N},
\end{equation}
and the fidelity  compared to the target ground state 
\be
f\bigr|_{(t = t_R)} =|\langle \psi_{\rm target} | \psi_{\rm final} \rangle|.
\ee
We also study the time-evolution of the local density 
\be
n_\sigma(x,t) = \langle \psi(t) | \hat n_{\sigma}(x) | \psi(t)\rangle.
\ee
In the following results we will report only the total density per grid point $n(x) = n_\uparrow(x) + n_\downarrow(x)$ since no local magnetization effects have been observed.
Additionally, we compute the local density integrated over one optical lattice unit cell 
\be
\overline n(i) = \sum_{k=1}^{N_{\rm discr}}n(x)\Bigr|_{x=(i+k-1)\,a}
\ee
which simplifies the analysis in terms of the effective lattice model, e.g. one expects $\overline n(i) =1$ in the Mott regime and  $\overline n(i) = 2$ in the band insulating regime.

 In order to reduce non-adiabticities we will propose an improved loading schedule that dynamically changes one or more parameters of the Hamiltonian, in addition to the lattice depth in the time dependent Hamiltonian. The first protocol follows the approach of Ref. \cite{dolfi2015} to dynamically reshape the trapping potential, by linearly modulating the trap frequency $\omega$. Starting with an initial value $\omega_i$, we increase $\omega$ linearly during the lattice loading to reach the desired target value $\omega_f$ at the end of ramp time.  At time $t$ the trap frequency is given by
\be\label{eq:tune_w}
\omega(t) = \omega_i + (\omega_f - \omega_i)\frac{t}{t_R}.
\ee
We perform simulations with different values of $w_i$ and different ramp times to study the scaling behavior. Our results for the improvements observed with this protocol are shown in sections \ref{sec:MI} and \ref{sec:MI_metal}.

Alternatively, we continuously tune the interaction strength during the time evolution, which is  more easily done in experiments via Feshbach resonances~\cite{esslinger, chin}. Since the density distribution of the initial state is found to be too narrow compared to the target state,  we initially use a stronger interaction $g_i$ to broaden the atomic cloud. $g(t)$ is then linearly reduced to its target value $g_f$:
\be\label{eq:tune_c}
g(t) = g_i - (g_i - g_f )\frac{t}{t_R}.
\ee

\subsection{Metallic target state}\label{sec:metallic}

\begin{figure}
\include{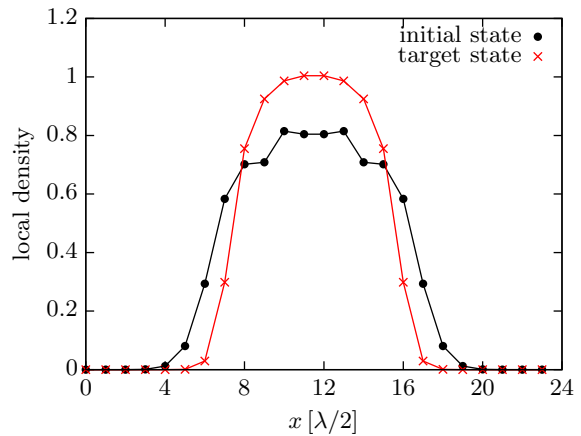}
\caption{Local density distribution of the initial state $(V_i=0)$ and the metallic target state $(V_f = V_0 = 8 E_r)$ integrated over each unit cell.}
\label{fig:metallic_states}
\end{figure}

We begin by studying the metallic target  states, where the fermions are delocalized over the lattice. Such a state is observed for a small number of particles $N<L_{\rm eff}$ and with a weak contact interaction. We simulated a chain  of $N = 8$ particles \textit{i.e.} $N_{\uparrow} = N_{\downarrow} = 4$. The interaction strength was chosen to be $g =0.2\, E_r \lambda / 2$ along with a trap frequency of $\omega = 0.1\, (\hbar/E_r)^{-1}$. The local density of this target state and the corresponding initial state without lattice potential are shown in Fig. \ref{fig:metallic_states}.

\begin{figure}[t]
\include{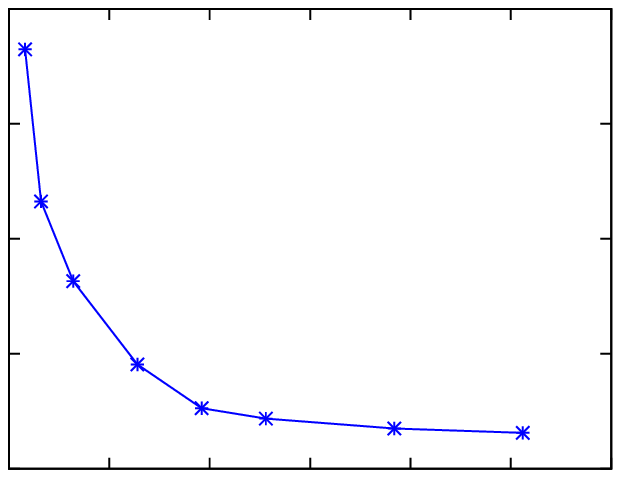}
\vspace{3mm}
\include{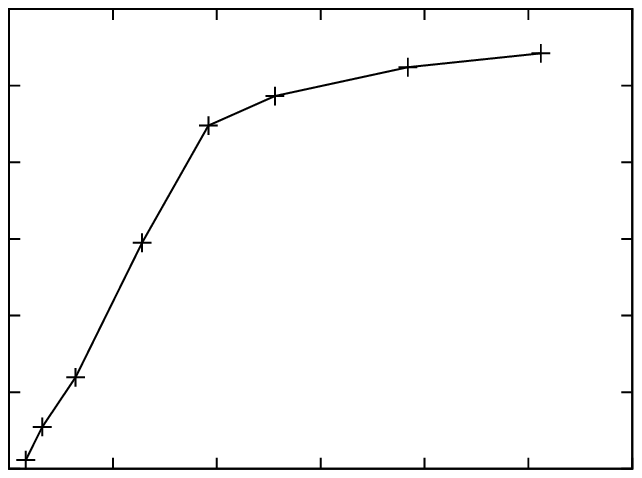}
\caption{Dependence on the ramp time of (a) excess energy and (b) fidelity for a metallic target state.}
\label{fig:metallic_heat_ov}
\end{figure}

Our simulation results, shown in Fig. \ref{fig:metallic_heat_ov}, indicate that it is possible to reach a final state fairly close to the target state just by slow loading the lattice. For $t_R = 256\, \hbar/E_r$, we observe a fidelity of more than $94\%$ and reduce the heating by a factor of 10 compared to shorter ramp times $t_R = 16\, \hbar/E_r$.

Intuitively one might think of the metallic state as a gapless state, with continuous excitations in momentum space. Hence this lattice loading could be extremely capable of populating low-lying excited states and at the risk of generating a lot of excess energy. However, given the finite system size $L_{\rm eff}$ originating from the harmonic confinement, there is always a finite gap that drastically reduces the excitations. 

\begin{figure}[t]
\include{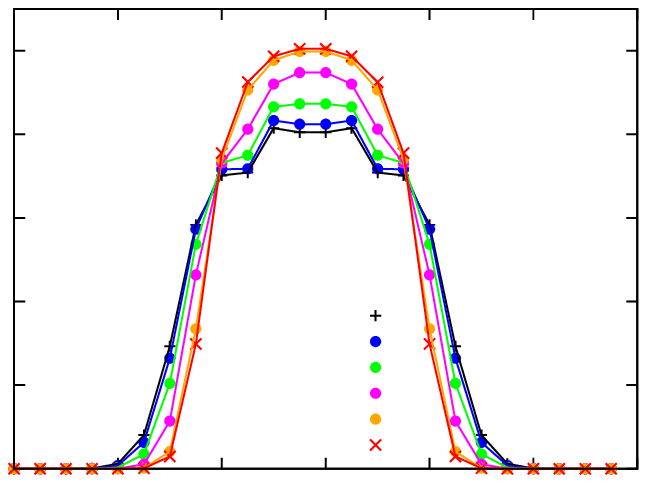}
\caption{Evolution of the density profile during the ramp up for ramp time $t_R =256\, \hbar / E_r$ for a metallic target state.}
\label{fig:metallic_ev}
\end{figure}

Additionally, the similarity of the density profiles of the initial and the target states allows the system to evolve almost without any defect. This is illustrated in the Fig.~\ref{fig:metallic_ev} showing snapshots of the density profile that have been collected at several time steps during the longest lattice loading up to  $t_R = 256\, \hbar / E_r$. The final state (orange curve) shows only very minimal deviations from the target state (red curve).

%%%%%%%%%%%%%%%%%%%%%%%%%%%%%%%%%%%%%%%%%%%%%%%%%%%%%%%%%%%%%%%%%%%%%%%%%%%%%%%%%%%%%%%%%%%%%%%%%%%%%%%%%
\subsection{Central Band Insulator}\label{sec:MI_BI}
Next we examine a target state that exhibits the co-existence of two phases: a central band insulating regime flanked by Mott insulating regions. Both these phases are incompressible and characterized by integer values of the average local density per unit cell. The Mott phase has one particle per lattice site, while the band insulator has an occupancy of two particles per site.

Here we consider a chain with a particle number $N$ close to (but less than) twice the effective system size $L_{\rm eff}$. Specifically, we choose $N=20$ particles ($N_{\uparrow} = N_{\downarrow} = 10$) with interaction strength $g= E_r \lambda / 2$ and trap frequency $\omega= 0.3\, (\hbar/E_r)^{-1}$.

\begin{figure}
\include{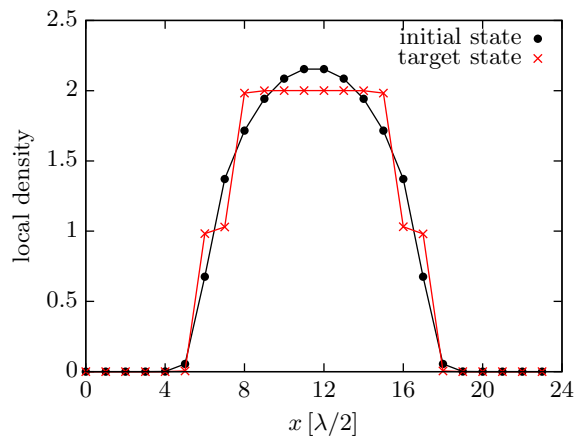}
\caption{Local density distribution of the initial state $(V_i=0)$ and the Mott insulator with bulk bad insulator target state $(V_f = V_0 = 8 E_r)$  integrated over each unit cell.}
\label{fig:mi_bi_states}
\end{figure}

The integrated local density distributions for the initial and target states are shown in Fig. \ref{fig:mi_bi_states}. In the target state the bulk of the system shows a band insulator phase while the edges are in the Mott insulator phase.

Note that again the density distributions of the initial state and the target state resemble each other in two important ways, namely the spatial spread of the system along with the peak value and its position in the density profile. This is the main reason why we notice that such a system does not incur significant heating if the lattice loading is done sufficiently slowly.

\begin{figure}
\include{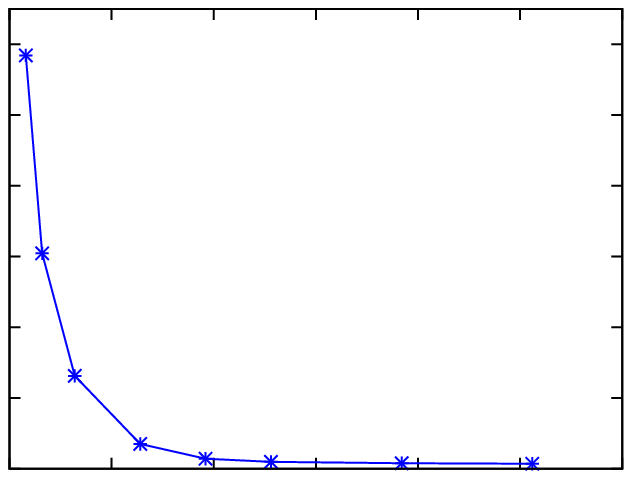}
\vspace{3mm}
\include{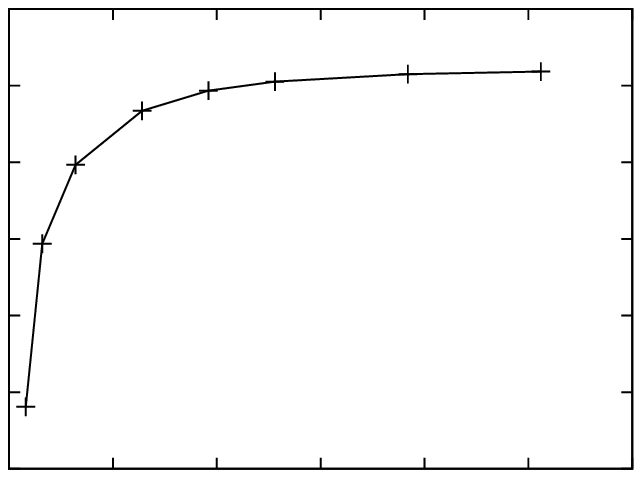}
\caption{Dependence of the (a) excess energy and (b) fidelity on ramp time for the case of a central band insulator. }
\label{fig:mi_bi_heat_ov}
\end{figure}

\begin{figure}
\include{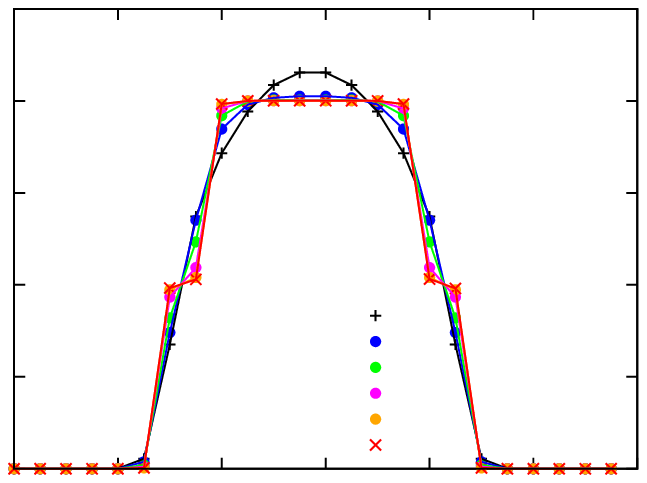}
\caption{Evolution of the density profile during the ramp up for ramp time $t_R =256\, \hbar / E_r$ for the case of a central band insulator. }
\label{fig:mi_bi_ev}
\end{figure}

The heating and fidelity as a function of the ramp time are shown in Fig. \ref{fig:mi_bi_heat_ov}(a) and (b) respectively. The fidelity grows rather rapidly initially upon increasing the loading time but then tends to saturate more or less around 90\%. For the shortest ramp time, the final density distribution is far from that of the target state. The proximity to the target state increases rapidly for the first few ramp times that we considered due to rapid changes in the density profile during the loading. Thereafter, the final state matches the target state to a good degree and further slowing down loading only brings about slight modifications in the density profile. This leads to a saturation of the fidelity. The evolution of the local density profile for the central band insulator state shown in Fig. \ref{fig:mi_bi_ev} for ramp time $t_R = 256 \, \hbar/E_r$ confirms a good match between the final and target state.

Note that although this target state is inherently incompressible as opposed to the metallic state studied in Section~\ref{sec:metallic}, it still does not suffer from adverse heating effects. This is a manifestation of the fact that the density distribution of the initial state that is in close qualitative correspondence in terms of $L_{\rm eff}$ and peak value, allowing an appropriate redistribution of particles during the ramp to reach the desired target state.

%%%%%%%%%%%%%%%%%%%%%%%%%%%%%%%%%%%%%%%%%%%%%%%%%%%%%%%%%%%%%%%%%%%%%%%%%%%%%%%%%%%%%%%%%%%%%%%%%%%%%%%%%
\subsection{Mott insulator}\label{sec:MI}

\begin{figure}[b]
\include{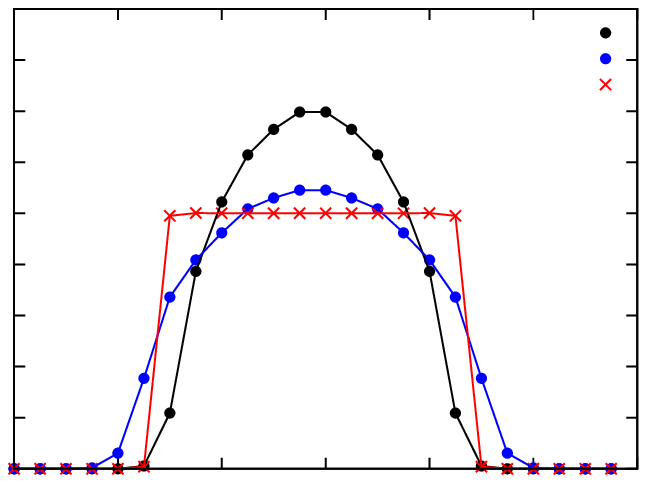}
\caption{Local density distribution of the initial state $(V_i=0)$ and the Mott insulator target state $(V_f = V_0 = 8 E_r)$ integrated over each unit cell. The blue curve shows the density distribution of the optimal state when linearly modulating the trap frequency $\omega$. }
\label{fig:mott_states}
\end{figure}

Our next choice of target state is a Mott insulator in the center of the trap. We choose $N=12$ particles ($N_{\uparrow} = N_{\downarrow} = 6$) with interaction strength $g= 2\, E_r \lambda / 2$ and trap frequency $\omega= 0.25\, (\hbar/E_r)^{-1}$. The local density profile of the initial state and the target state are shown in Fig. \ref{fig:mott_states}.

\begin{figure}
\include{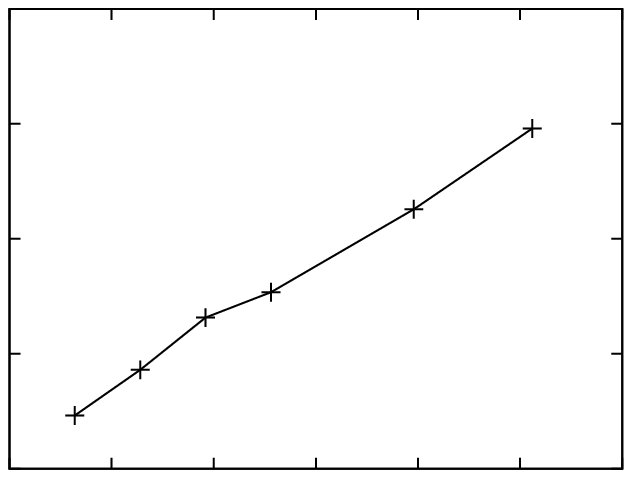}
\vspace{3mm}
\include{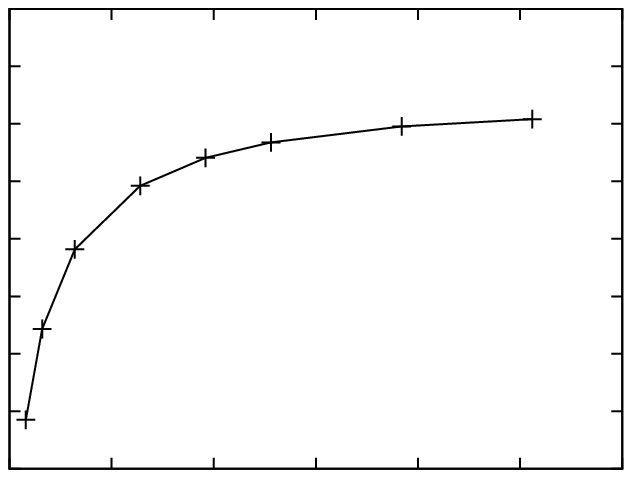}
\caption{Dependence of the fidelity on ramp time for a Mott insulator target state in the (a) trapped case and (b) homogenous case.}
\label{fig:ov_fixed_w_mott}
\end{figure}

For this target state we observe that increasing ramp times does not lead to high fidelity. Even for the longest ramp time that we considered $(t_R = 256\, \hbar / E_r)$, the highest fidelity we achieve is only about $15 \%$, as shown in Fig.~\ref{fig:ov_fixed_w_mott}(a). Though by further increasing the ramp time, we should be able to get a better fidelity but possibly not an impressive increase. The slow increase of fidelity with ramp time is a clear indication that simulating the lattice loading with a finite ramp time is not the main cause of heating in the system. A fermionic Mott insulator state in a homogenous system (without a trapping potential) does not suffer from strong defects and both the excess energy and the fidelity scale well up to, for instance, a fidelity of $80\%$ for ramp time $t_R = 256\, \hbar / E_r$ as shown in Fig. \ref{fig:ov_fixed_w_mott}(b). This is far from the value observed for the trapped Mott insulator, which hints at the harmonic trap being a plausible source of heating.

\begin{figure}
\include{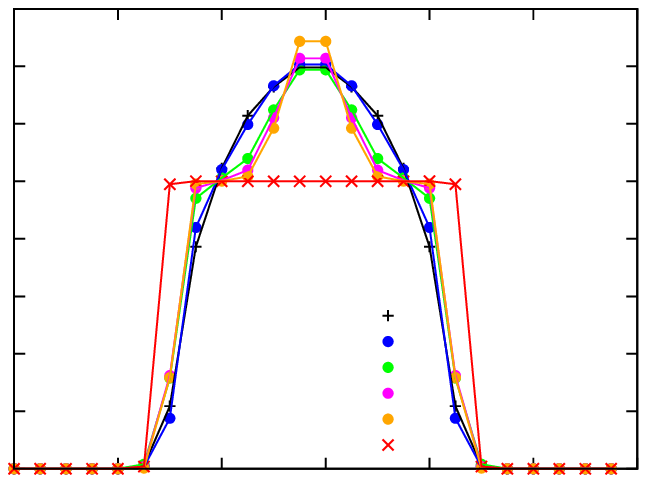}
\vspace{3mm}
\include{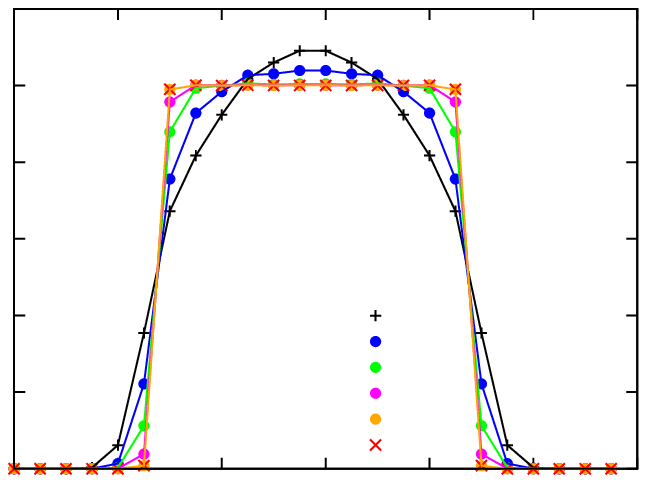}
\caption{Evolution of the density profile during the ramp up for a Mott insulator target state with ramp time $t_R =256\, \hbar / E_r$ (a) without tuning the trap frequency (b) with linear modulation of the frequency, for the optimal value of initial frequency $(w_i =0.16\, (\hbar/E_r)^{-1}$). The black line corresponds to the target state.}
\label{fig:mott_ev}
\end{figure}

In Fig. \ref{fig:mott_ev}(a) we show the evolution of density profile during lattice loading. This plot reveals that the loading process is not able to distribute particles in the desired way, thus deviating significantly from the target state. As it is evident from the figure, the evolution tends to keep the particles close to the trap centre and this peak remains until the end of the ramp time. 

From our previous analyses in sections \ref{sec:metallic} and \ref{sec:MI_BI}, we understand that a a qualitative match between the density of the initial and target states is imperative to avoid strong density defects during lattice loading, which can be achieved by dynamically changing system parameters during loading.

\begin{figure}
\include{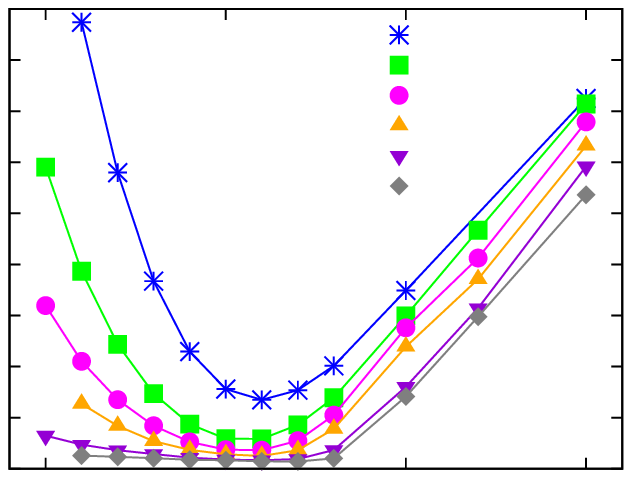}
\vspace{3mm}
\include{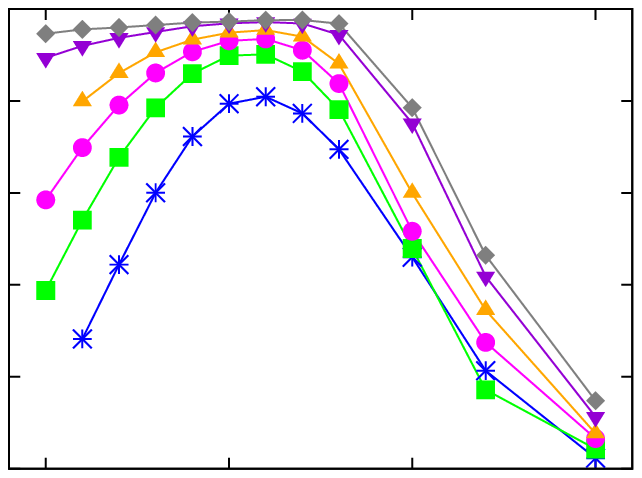}
\caption{Variation of (a) excess energy and (b) fidelity as a function of the initial frequency for a Mott insulator target state. The different colors correspond to different ramp times. }
\label{fig:mott_heat_ov}
\end{figure}

We first linearly modulate the trap frequency. Fig. \ref{fig:mott_heat_ov} shows the excess energy  and fidelity scanning different initial trap frequencies $\omega_i$.  We observe a huge improvement in the fidelity and a sizeable decrease in the excess energy, compared to the case when the trap frequency is fixed during the lattice loading, which corresponds to the right-most point in the plots.

From the shape of the curves we can identify three different scaling behaviors. Close to the target $\omega_f$ the observables do not show any appreciable variation and the results are always significantly different than those of the target state, while wide initial traps (low $\omega_i$) reach the target state, but this process scales slowly. An \emph{optimal} and fast scaling is observed for intermediate  $\omega_i$. We can identify an optimal initial state marked by a maximum in the fidelity curve, which happens to be at $w_i = 0.16 (\hbar/E_r)^{-1}$ for our particular simulation.

The local density distribution of the optimal initial state is shown by the blue curve in Fig. \ref{fig:mott_states}. 
The maximum fidelity achieved for $t_R = 256\, \hbar / E_r$ is almost $98 \%$ and the heating is reduced by a factor 50.

 In Fig. \ref{fig:mott_ev}(b) we show the evolution of the density profile for this optimal state during the lattice loading for $t_R = 256\, \hbar / E_r$. This is in stark contrast with the evolution plot of Fig. \ref{fig:mott_ev}(a) where the trap frequency remained constant during the entire process of loading. From $t=0$ to already at time $t_1=t_R/4$, the density profile is changed drastically when the trap frequency is modulated which was not the case earlier. Also the time evolved state at time $t_1$ is nearly a Mott state in the trap centre whereas it had a more metallic nature in the previous case. At the end of ramp time, the density profile of the final state is almost exactly that of the target state, corresponding to an overlap of almost $98\%$ (as can be seen in fig. \ref{fig:mott_heat_ov}(b)).

We can thus conclude that density redistribution is the main cause of heating. By tuning the trap frequency during the lattice loading we are able to distribute the particles more efficiently, thus we observe a remarkable jump in the fidelity. An optimal initial state is the one with a considerable matching to the target density profiles.

%%%%%%%%%%%%%%%%%%%%%%%%%%%%%%%%%%%%%%%%%%%%%%%%%%%%%%%%%%%%%%%%%%%%%%%%%%%%%%%%%%%%%%%%%%%%%%%%%%%%%%%%%
\subsection{Mott insulator with a metallic core}\label{sec:MI_metal}

\begin{figure}
\include{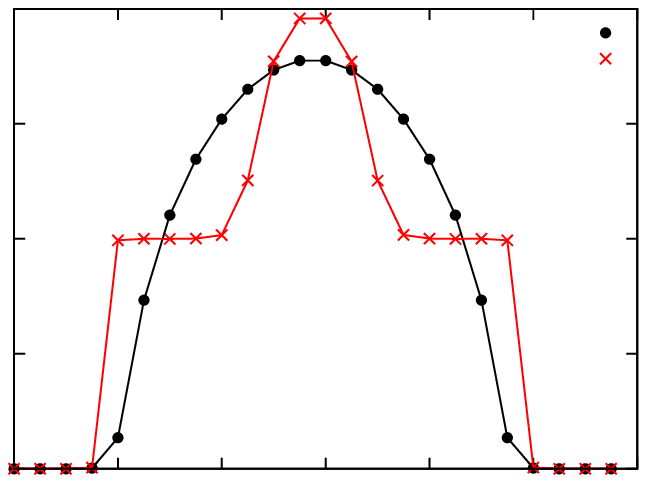}
\caption{Local density distribution of the initial state $(V_i=0)$ and the Mott insulator with a metallic core target state $(V_f = V_0 = 8 E_r)$ integrated over each unit cell.}
\label{fig:mi_metal_states}
\end{figure}

The last target state we investigate is the Mott insulator state with a metallic core, as shown in Fig. \ref{fig:mi_metal_states}.  
 Our simulations are done with $N=20$ particles ($N_{\uparrow} = N_{\downarrow} = 10$) with interaction strength $g= 3\, E_r \lambda / 2$ and trap frequency $\omega= 0.25\, (\hbar/E_r)^{-1}$.

\begin{figure}
\include{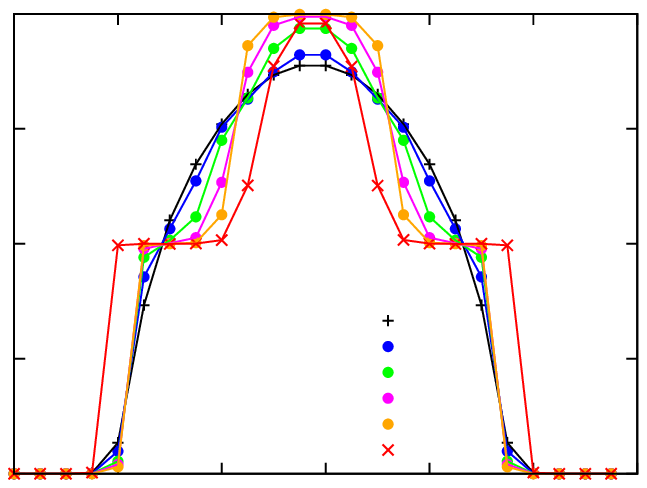}
\vspace{3mm}
\include{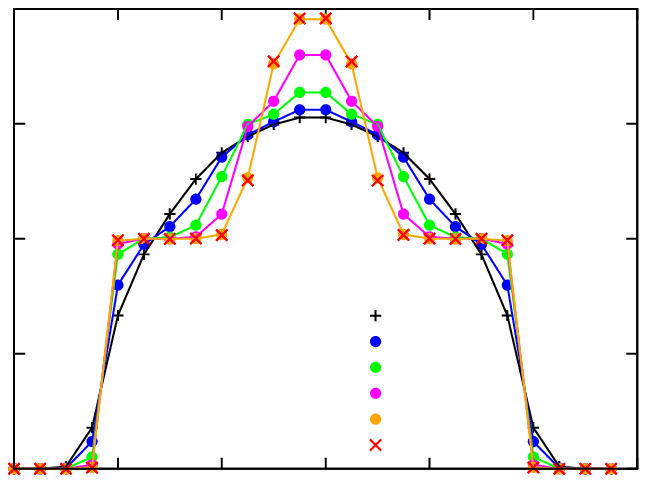}
\caption{Evolution of the density profile during the ramp up for the case of a Mott insulator with a metallic core at ramp time $t_R =256\, \hbar / E_r$ (a) without tuning the trap frequency (b) with linear modulation of the frequency, for the optimal value of initial frequency $(w_i =0.19 \, (\hbar/E_r)^{-1}$). The red line corresponds to the target state.}
\label{fig:mi_metal_ev}
\end{figure}
Ramping up the lattice potential shows severe heating and the fidelity remains less than 2\% even for the longest ramp times simulated $(t_R = 256\, \hbar / E_r)$. We again  track the evolution of the local density during the ramp up, as we show in Fig. \ref{fig:mi_metal_ev}(a). It reveals that the final state has a core with a substantial number of sites having a local density of two particles per unit cell, tending towards a band insulator core rather than a metallic one.

\begin {figure}
% GNUPLOT: LaTeX picture with Postscript
\begingroup
  \makeatletter
  \providecommand\color[2][]{%
    \GenericError{(gnuplot) \space\space\space\@spaces}{%
      Package color not loaded in conjunction with
      terminal option `colourtext'%
    }{See the gnuplot documentation for explanation.%
    }{Either use 'blacktext' in gnuplot or load the package
      color.sty in LaTeX.}%
    \renewcommand\color[2][]{}%
  }%
  \providecommand\includegraphics[2][]{%
    \GenericError{(gnuplot) \space\space\space\@spaces}{%
      Package graphicx or graphics not loaded%
    }{See the gnuplot documentation for explanation.%
    }{The gnuplot epslatex terminal needs graphicx.sty or graphics.sty.}%
    \renewcommand\includegraphics[2][]{}%
  }%
  \providecommand\rotatebox[2]{#2}%
  \@ifundefined{ifGPcolor}{%
    \newif\ifGPcolor
    \GPcolortrue
  }{}%
  \@ifundefined{ifGPblacktext}{%
    \newif\ifGPblacktext
    \GPblacktexttrue
  }{}%
  % define a \g@addto@macro without @ in the name:
  \let\gplgaddtomacro\g@addto@macro
  % define empty templates for all commands taking text:
  \gdef\gplbacktext{}%
  \gdef\gplfronttext{}%
  \makeatother
  \ifGPblacktext
    % no textcolor at all
    \def\colorrgb#1{}%
    \def\colorgray#1{}%
  \else
    % gray or color?
    \ifGPcolor
      \def\colorrgb#1{\color[rgb]{#1}}%
      \def\colorgray#1{\color[gray]{#1}}%
      \expandafter\def\csname LTw\endcsname{\color{white}}%
      \expandafter\def\csname LTb\endcsname{\color{black}}%
      \expandafter\def\csname LTa\endcsname{\color{black}}%
      \expandafter\def\csname LT0\endcsname{\color[rgb]{1,0,0}}%
      \expandafter\def\csname LT1\endcsname{\color[rgb]{0,1,0}}%
      \expandafter\def\csname LT2\endcsname{\color[rgb]{0,0,1}}%
      \expandafter\def\csname LT3\endcsname{\color[rgb]{1,0,1}}%
      \expandafter\def\csname LT4\endcsname{\color[rgb]{0,1,1}}%
      \expandafter\def\csname LT5\endcsname{\color[rgb]{1,1,0}}%
      \expandafter\def\csname LT6\endcsname{\color[rgb]{0,0,0}}%
      \expandafter\def\csname LT7\endcsname{\color[rgb]{1,0.3,0}}%
      \expandafter\def\csname LT8\endcsname{\color[rgb]{0.5,0.5,0.5}}%
    \else
      % gray
      \def\colorrgb#1{\color{black}}%
      \def\colorgray#1{\color[gray]{#1}}%
      \expandafter\def\csname LTw\endcsname{\color{white}}%
      \expandafter\def\csname LTb\endcsname{\color{black}}%
      \expandafter\def\csname LTa\endcsname{\color{black}}%
      \expandafter\def\csname LT0\endcsname{\color{black}}%
      \expandafter\def\csname LT1\endcsname{\color{black}}%
      \expandafter\def\csname LT2\endcsname{\color{black}}%
      \expandafter\def\csname LT3\endcsname{\color{black}}%
      \expandafter\def\csname LT4\endcsname{\color{black}}%
      \expandafter\def\csname LT5\endcsname{\color{black}}%
      \expandafter\def\csname LT6\endcsname{\color{black}}%
      \expandafter\def\csname LT7\endcsname{\color{black}}%
      \expandafter\def\csname LT8\endcsname{\color{black}}%
    \fi
  \fi
    \setlength{\unitlength}{0.0500bp}%
    \ifx\gptboxheight\undefined%
      \newlength{\gptboxheight}%
      \newlength{\gptboxwidth}%
      \newsavebox{\gptboxtext}%
    \fi%
    \setlength{\fboxrule}{0.5pt}%
    \setlength{\fboxsep}{1pt}%
\begin{picture}(4320.00,3240.00)%
    \gplgaddtomacro\gplbacktext{%
      \csname LTb\endcsname%
      \put(610,470){\makebox(0,0)[r]{\strut{}$0$}}%
      \put(610,1046){\makebox(0,0)[r]{\strut{}$0.05$}}%
      \put(610,1622){\makebox(0,0)[r]{\strut{}$0.1$}}%
      \put(610,2198){\makebox(0,0)[r]{\strut{}$0.15$}}%
      \put(610,2773){\makebox(0,0)[r]{\strut{}$0.2$}}%
      \put(874,320){\makebox(0,0){\strut{}$0.1$}}%
      \put(1894,320){\makebox(0,0){\strut{}$0.15$}}%
      \put(2915,320){\makebox(0,0){\strut{}$0.2$}}%
      \put(3935,320){\makebox(0,0){\strut{}$0.25$}}%
      \put(874,2773){\makebox(0,0){\strut{}(a)}}%
    }%
    \gplgaddtomacro\gplfronttext{%
      \csname LTb\endcsname%
      \put(80,1794){\rotatebox{-270}{\makebox(0,0){\strut{}$q(t_R)/E_r$}}}%
      \put(2404,70){\makebox(0,0){\strut{}$\omega_i\, \big[(\hbar/E_r)^{-1}\big]$}}%
      \csname LTb\endcsname%
      \put(3612,2969){\makebox(0,0)[r]{\strut{}$t_R = 16 \, \hbar/E_r$}}%
      \csname LTb\endcsname%
      \put(3612,2795){\makebox(0,0)[r]{\strut{}$t_R = 32 \, \hbar/E_r$}}%
      \csname LTb\endcsname%
      \put(3612,2621){\makebox(0,0)[r]{\strut{}$t_R = 64 \, \hbar/E_r$}}%
      \csname LTb\endcsname%
      \put(3612,2447){\makebox(0,0)[r]{\strut{}$t_R = 96 \, \hbar/E_r$}}%
      \csname LTb\endcsname%
      \put(3612,2273){\makebox(0,0)[r]{\strut{}$t_R = 128 \, \hbar/E_r$}}%
      \csname LTb\endcsname%
      \put(3612,2099){\makebox(0,0)[r]{\strut{}$t_R = 192 \, \hbar/E_r$}}%
      \csname LTb\endcsname%
      \put(3612,1925){\makebox(0,0)[r]{\strut{}$t_R = 256 \, \hbar/E_r$}}%
    }%
    \gplbacktext
    \put(0,0){\includegraphics{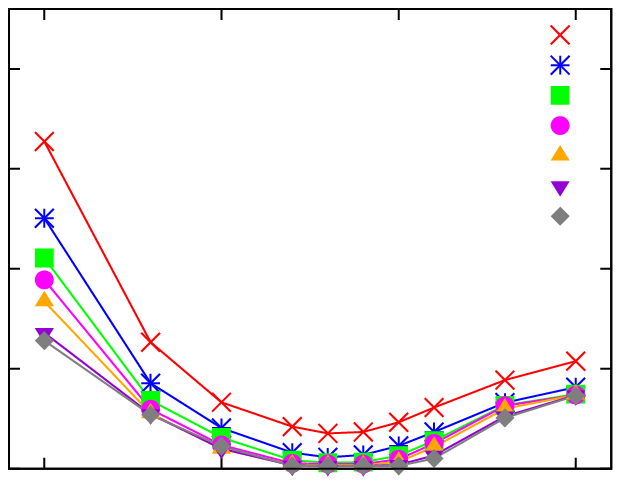}}%
    \gplfronttext
  \end{picture}%
\endgroup

\vspace{3mm}
% GNUPLOT: LaTeX picture with Postscript
\begingroup
  \makeatletter
  \providecommand\color[2][]{%
    \GenericError{(gnuplot) \space\space\space\@spaces}{%
      Package color not loaded in conjunction with
      terminal option `colourtext'%
    }{See the gnuplot documentation for explanation.%
    }{Either use 'blacktext' in gnuplot or load the package
      color.sty in LaTeX.}%
    \renewcommand\color[2][]{}%
  }%
  \providecommand\includegraphics[2][]{%
    \GenericError{(gnuplot) \space\space\space\@spaces}{%
      Package graphicx or graphics not loaded%
    }{See the gnuplot documentation for explanation.%
    }{The gnuplot epslatex terminal needs graphicx.sty or graphics.sty.}%
    \renewcommand\includegraphics[2][]{}%
  }%
  \providecommand\rotatebox[2]{#2}%
  \@ifundefined{ifGPcolor}{%
    \newif\ifGPcolor
    \GPcolortrue
  }{}%
  \@ifundefined{ifGPblacktext}{%
    \newif\ifGPblacktext
    \GPblacktexttrue
  }{}%
  % define a \g@addto@macro without @ in the name:
  \let\gplgaddtomacro\g@addto@macro
  % define empty templates for all commands taking text:
  \gdef\gplbacktext{}%
  \gdef\gplfronttext{}%
  \makeatother
  \ifGPblacktext
    % no textcolor at all
    \def\colorrgb#1{}%
    \def\colorgray#1{}%
  \else
    % gray or color?
    \ifGPcolor
      \def\colorrgb#1{\color[rgb]{#1}}%
      \def\colorgray#1{\color[gray]{#1}}%
      \expandafter\def\csname LTw\endcsname{\color{white}}%
      \expandafter\def\csname LTb\endcsname{\color{black}}%
      \expandafter\def\csname LTa\endcsname{\color{black}}%
      \expandafter\def\csname LT0\endcsname{\color[rgb]{1,0,0}}%
      \expandafter\def\csname LT1\endcsname{\color[rgb]{0,1,0}}%
      \expandafter\def\csname LT2\endcsname{\color[rgb]{0,0,1}}%
      \expandafter\def\csname LT3\endcsname{\color[rgb]{1,0,1}}%
      \expandafter\def\csname LT4\endcsname{\color[rgb]{0,1,1}}%
      \expandafter\def\csname LT5\endcsname{\color[rgb]{1,1,0}}%
      \expandafter\def\csname LT6\endcsname{\color[rgb]{0,0,0}}%
      \expandafter\def\csname LT7\endcsname{\color[rgb]{1,0.3,0}}%
      \expandafter\def\csname LT8\endcsname{\color[rgb]{0.5,0.5,0.5}}%
    \else
      % gray
      \def\colorrgb#1{\color{black}}%
      \def\colorgray#1{\color[gray]{#1}}%
      \expandafter\def\csname LTw\endcsname{\color{white}}%
      \expandafter\def\csname LTb\endcsname{\color{black}}%
      \expandafter\def\csname LTa\endcsname{\color{black}}%
      \expandafter\def\csname LT0\endcsname{\color{black}}%
      \expandafter\def\csname LT1\endcsname{\color{black}}%
      \expandafter\def\csname LT2\endcsname{\color{black}}%
      \expandafter\def\csname LT3\endcsname{\color{black}}%
      \expandafter\def\csname LT4\endcsname{\color{black}}%
      \expandafter\def\csname LT5\endcsname{\color{black}}%
      \expandafter\def\csname LT6\endcsname{\color{black}}%
      \expandafter\def\csname LT7\endcsname{\color{black}}%
      \expandafter\def\csname LT8\endcsname{\color{black}}%
    \fi
  \fi
    \setlength{\unitlength}{0.0500bp}%
    \ifx\gptboxheight\undefined%
      \newlength{\gptboxheight}%
      \newlength{\gptboxwidth}%
      \newsavebox{\gptboxtext}%
    \fi%
    \setlength{\fboxrule}{0.5pt}%
    \setlength{\fboxsep}{1pt}%
\begin{picture}(4320.00,3240.00)%
    \gplgaddtomacro\gplbacktext{%
      \csname LTb\endcsname%
      \put(550,470){\makebox(0,0)[r]{\strut{}$0$}}%
      \put(550,1000){\makebox(0,0)[r]{\strut{}$0.2$}}%
      \put(550,1530){\makebox(0,0)[r]{\strut{}$0.4$}}%
      \put(550,2059){\makebox(0,0)[r]{\strut{}$0.6$}}%
      \put(550,2589){\makebox(0,0)[r]{\strut{}$0.8$}}%
      \put(550,3119){\makebox(0,0)[r]{\strut{}$1$}}%
      \put(818,320){\makebox(0,0){\strut{}$0.1$}}%
      \put(1856,320){\makebox(0,0){\strut{}$0.15$}}%
      \put(2893,320){\makebox(0,0){\strut{}$0.2$}}%
      \put(3931,320){\makebox(0,0){\strut{}$0.25$}}%
      \put(818,2722){\makebox(0,0){\strut{}(b)}}%
    }%
    \gplgaddtomacro\gplfronttext{%
      \csname LTb\endcsname%
      \put(80,1794){\rotatebox{-270}{\makebox(0,0){\strut{}$|\langle\psi_{\rm target}|\psi_{\rm final}\rangle|$}}}%
      \put(2374,70){\makebox(0,0){\strut{}$\omega_i\, \big[(\hbar/E_r)^{-1}\big]$}}%
    }%
    \gplbacktext
    \put(0,0){\includegraphics{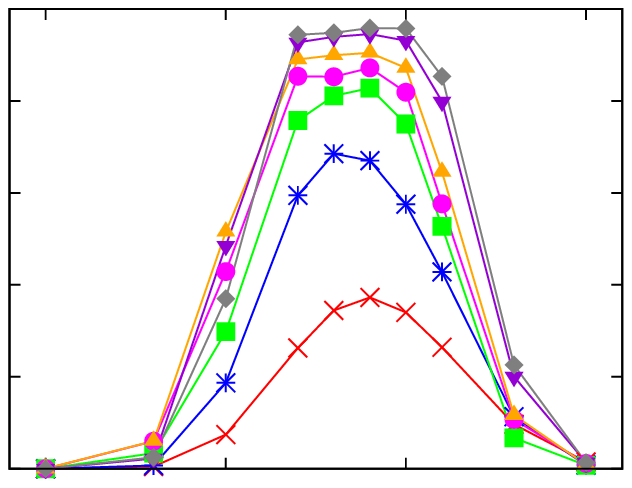}}%
    \gplfronttext
  \end{picture}%
\endgroup

\caption{Variation of (a) excess energy and (b) fidelity as a function of the initial frequency for the case of a Mott insulator with a metallic core. The different colors correspond to different ramp times.}
\label{fig:mi_metal_heat_ov_w}
\end {figure}

In order to reduce these defects we again dynamically change model parameters during loading.  Varying the trap frequency improves the fidelity compared to the target state to about $96\%$ at ramp time $t_R = 256\, \hbar / E_r$ for  $\omega_i = 0.19\, (\hbar/E_r)^{-1}$. The excess energy is also reduced significantly. Figure \ref{fig:mi_metal_heat_ov_w} shows the heating and fidelity under trap modulation for different ramp times. The optimal initial state has two characteristic features: it mimics the target state in the extent of the density distribution and secondly it lowers the peak value of the density.

\begin {figure}[t]
% GNUPLOT: LaTeX picture with Postscript
\begingroup
  \makeatletter
  \providecommand\color[2][]{%
    \GenericError{(gnuplot) \space\space\space\@spaces}{%
      Package color not loaded in conjunction with
      terminal option `colourtext'%
    }{See the gnuplot documentation for explanation.%
    }{Either use 'blacktext' in gnuplot or load the package
      color.sty in LaTeX.}%
    \renewcommand\color[2][]{}%
  }%
  \providecommand\includegraphics[2][]{%
    \GenericError{(gnuplot) \space\space\space\@spaces}{%
      Package graphicx or graphics not loaded%
    }{See the gnuplot documentation for explanation.%
    }{The gnuplot epslatex terminal needs graphicx.sty or graphics.sty.}%
    \renewcommand\includegraphics[2][]{}%
  }%
  \providecommand\rotatebox[2]{#2}%
  \@ifundefined{ifGPcolor}{%
    \newif\ifGPcolor
    \GPcolortrue
  }{}%
  \@ifundefined{ifGPblacktext}{%
    \newif\ifGPblacktext
    \GPblacktexttrue
  }{}%
  % define a \g@addto@macro without @ in the name:
  \let\gplgaddtomacro\g@addto@macro
  % define empty templates for all commands taking text:
  \gdef\gplbacktext{}%
  \gdef\gplfronttext{}%
  \makeatother
  \ifGPblacktext
    % no textcolor at all
    \def\colorrgb#1{}%
    \def\colorgray#1{}%
  \else
    % gray or color?
    \ifGPcolor
      \def\colorrgb#1{\color[rgb]{#1}}%
      \def\colorgray#1{\color[gray]{#1}}%
      \expandafter\def\csname LTw\endcsname{\color{white}}%
      \expandafter\def\csname LTb\endcsname{\color{black}}%
      \expandafter\def\csname LTa\endcsname{\color{black}}%
      \expandafter\def\csname LT0\endcsname{\color[rgb]{1,0,0}}%
      \expandafter\def\csname LT1\endcsname{\color[rgb]{0,1,0}}%
      \expandafter\def\csname LT2\endcsname{\color[rgb]{0,0,1}}%
      \expandafter\def\csname LT3\endcsname{\color[rgb]{1,0,1}}%
      \expandafter\def\csname LT4\endcsname{\color[rgb]{0,1,1}}%
      \expandafter\def\csname LT5\endcsname{\color[rgb]{1,1,0}}%
      \expandafter\def\csname LT6\endcsname{\color[rgb]{0,0,0}}%
      \expandafter\def\csname LT7\endcsname{\color[rgb]{1,0.3,0}}%
      \expandafter\def\csname LT8\endcsname{\color[rgb]{0.5,0.5,0.5}}%
    \else
      % gray
      \def\colorrgb#1{\color{black}}%
      \def\colorgray#1{\color[gray]{#1}}%
      \expandafter\def\csname LTw\endcsname{\color{white}}%
      \expandafter\def\csname LTb\endcsname{\color{black}}%
      \expandafter\def\csname LTa\endcsname{\color{black}}%
      \expandafter\def\csname LT0\endcsname{\color{black}}%
      \expandafter\def\csname LT1\endcsname{\color{black}}%
      \expandafter\def\csname LT2\endcsname{\color{black}}%
      \expandafter\def\csname LT3\endcsname{\color{black}}%
      \expandafter\def\csname LT4\endcsname{\color{black}}%
      \expandafter\def\csname LT5\endcsname{\color{black}}%
      \expandafter\def\csname LT6\endcsname{\color{black}}%
      \expandafter\def\csname LT7\endcsname{\color{black}}%
      \expandafter\def\csname LT8\endcsname{\color{black}}%
    \fi
  \fi
    \setlength{\unitlength}{0.0500bp}%
    \ifx\gptboxheight\undefined%
      \newlength{\gptboxheight}%
      \newlength{\gptboxwidth}%
      \newsavebox{\gptboxtext}%
    \fi%
    \setlength{\fboxrule}{0.5pt}%
    \setlength{\fboxsep}{1pt}%
\begin{picture}(4320.00,3240.00)%
    \gplgaddtomacro\gplbacktext{%
      \csname LTb\endcsname%
      \put(490,470){\makebox(0,0)[r]{\strut{}$0$}}%
      \put(490,1000){\makebox(0,0)[r]{\strut{}$0.2$}}%
      \put(490,1530){\makebox(0,0)[r]{\strut{}$0.4$}}%
      \put(490,2059){\makebox(0,0)[r]{\strut{}$0.6$}}%
      \put(490,2589){\makebox(0,0)[r]{\strut{}$0.8$}}%
      \put(490,3119){\makebox(0,0)[r]{\strut{}$1$}}%
      \put(550,320){\makebox(0,0){\strut{}$3$}}%
      \put(1022,320){\makebox(0,0){\strut{}$5$}}%
      \put(1494,320){\makebox(0,0){\strut{}$7$}}%
      \put(1967,320){\makebox(0,0){\strut{}$9$}}%
      \put(2439,320){\makebox(0,0){\strut{}$11$}}%
      \put(2911,320){\makebox(0,0){\strut{}$13$}}%
      \put(3383,320){\makebox(0,0){\strut{}$15$}}%
      \put(3856,320){\makebox(0,0){\strut{}$17$}}%
    }%
    \gplgaddtomacro\gplfronttext{%
      \csname LTb\endcsname%
      \put(80,1794){\rotatebox{-270}{\makebox(0,0){\strut{}$|\langle\psi_{\rm target}|\psi_{\rm final}\rangle|$}}}%
      \put(2344,70){\makebox(0,0){\strut{}$g_i\, [E_r \lambda/2]$}}%
    }%
    \gplbacktext
    \put(0,0){\includegraphics{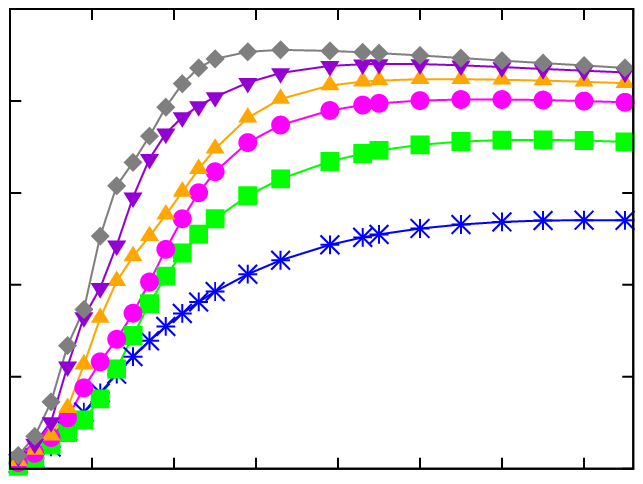}}%
    \gplfronttext
  \end{picture}%
\endgroup

\caption{Variation of the fidelity as a function of the interaction strength for the case of a Mott insulator with a metallic core. The different colors correspond to different ramp times. Color scheme remains same as in Fig. \ref{fig:mi_metal_heat_ov_w}.}
\label{fig:mi_metal_heat_ov_c}
\end {figure}

We next modify the interaction strength during loading according to Eq. (\ref{eq:tune_c}). This approach also provides a qualitative improvement of the fidelity.  Figure \ref{fig:mi_metal_heat_ov_c} shows the fidelity as a function of interaction strength for different ramp times. It is evident that this protocol leads to a more extended optimal regime where the fidelity is maximized. A similar pattern for the evolution of the local density with longest ramp time is observed as in Fig. \ref{fig:mi_metal_ev}(b) starting from initial states with interaction strength lying at the optimal plateau.

Yet another approach is to to combine the above two approaches. In such a scenario, we tune both the interaction strength and the trap frequency during the lattice loading. In our simulations we see that linearly modulating both trap frequency and interaction strength, starting from a good candidate state, efficiently distributes the particles during the lattice loading.

\begin {figure}
% GNUPLOT: LaTeX picture with Postscript
\begingroup
  \makeatletter
  \providecommand\color[2][]{%
    \GenericError{(gnuplot) \space\space\space\@spaces}{%
      Package color not loaded in conjunction with
      terminal option `colourtext'%
    }{See the gnuplot documentation for explanation.%
    }{Either use 'blacktext' in gnuplot or load the package
      color.sty in LaTeX.}%
    \renewcommand\color[2][]{}%
  }%
  \providecommand\includegraphics[2][]{%
    \GenericError{(gnuplot) \space\space\space\@spaces}{%
      Package graphicx or graphics not loaded%
    }{See the gnuplot documentation for explanation.%
    }{The gnuplot epslatex terminal needs graphicx.sty or graphics.sty.}%
    \renewcommand\includegraphics[2][]{}%
  }%
  \providecommand\rotatebox[2]{#2}%
  \@ifundefined{ifGPcolor}{%
    \newif\ifGPcolor
    \GPcolortrue
  }{}%
  \@ifundefined{ifGPblacktext}{%
    \newif\ifGPblacktext
    \GPblacktexttrue
  }{}%
  % define a \g@addto@macro without @ in the name:
  \let\gplgaddtomacro\g@addto@macro
  % define empty templates for all commands taking text:
  \gdef\gplbacktext{}%
  \gdef\gplfronttext{}%
  \makeatother
  \ifGPblacktext
    % no textcolor at all
    \def\colorrgb#1{}%
    \def\colorgray#1{}%
  \else
    % gray or color?
    \ifGPcolor
      \def\colorrgb#1{\color[rgb]{#1}}%
      \def\colorgray#1{\color[gray]{#1}}%
      \expandafter\def\csname LTw\endcsname{\color{white}}%
      \expandafter\def\csname LTb\endcsname{\color{black}}%
      \expandafter\def\csname LTa\endcsname{\color{black}}%
      \expandafter\def\csname LT0\endcsname{\color[rgb]{1,0,0}}%
      \expandafter\def\csname LT1\endcsname{\color[rgb]{0,1,0}}%
      \expandafter\def\csname LT2\endcsname{\color[rgb]{0,0,1}}%
      \expandafter\def\csname LT3\endcsname{\color[rgb]{1,0,1}}%
      \expandafter\def\csname LT4\endcsname{\color[rgb]{0,1,1}}%
      \expandafter\def\csname LT5\endcsname{\color[rgb]{1,1,0}}%
      \expandafter\def\csname LT6\endcsname{\color[rgb]{0,0,0}}%
      \expandafter\def\csname LT7\endcsname{\color[rgb]{1,0.3,0}}%
      \expandafter\def\csname LT8\endcsname{\color[rgb]{0.5,0.5,0.5}}%
    \else
      % gray
      \def\colorrgb#1{\color{black}}%
      \def\colorgray#1{\color[gray]{#1}}%
      \expandafter\def\csname LTw\endcsname{\color{white}}%
      \expandafter\def\csname LTb\endcsname{\color{black}}%
      \expandafter\def\csname LTa\endcsname{\color{black}}%
      \expandafter\def\csname LT0\endcsname{\color{black}}%
      \expandafter\def\csname LT1\endcsname{\color{black}}%
      \expandafter\def\csname LT2\endcsname{\color{black}}%
      \expandafter\def\csname LT3\endcsname{\color{black}}%
      \expandafter\def\csname LT4\endcsname{\color{black}}%
      \expandafter\def\csname LT5\endcsname{\color{black}}%
      \expandafter\def\csname LT6\endcsname{\color{black}}%
      \expandafter\def\csname LT7\endcsname{\color{black}}%
      \expandafter\def\csname LT8\endcsname{\color{black}}%
    \fi
  \fi
    \setlength{\unitlength}{0.0500bp}%
    \ifx\gptboxheight\undefined%
      \newlength{\gptboxheight}%
      \newlength{\gptboxwidth}%
      \newsavebox{\gptboxtext}%
    \fi%
    \setlength{\fboxrule}{0.5pt}%
    \setlength{\fboxsep}{1pt}%
\begin{picture}(4320.00,3240.00)%
    \gplgaddtomacro\gplbacktext{%
      \csname LTb\endcsname%
      \put(550,470){\makebox(0,0)[r]{\strut{}$0$}}%
      \put(550,1132){\makebox(0,0)[r]{\strut{}$0.5$}}%
      \put(550,1795){\makebox(0,0)[r]{\strut{}$1$}}%
      \put(550,2457){\makebox(0,0)[r]{\strut{}$1.5$}}%
      \put(550,3119){\makebox(0,0)[r]{\strut{}$2$}}%
      \put(610,320){\makebox(0,0){\strut{}$0$}}%
      \put(1175,320){\makebox(0,0){\strut{}$4$}}%
      \put(1739,320){\makebox(0,0){\strut{}$8$}}%
      \put(2304,320){\makebox(0,0){\strut{}$12$}}%
      \put(2869,320){\makebox(0,0){\strut{}$16$}}%
      \put(3433,320){\makebox(0,0){\strut{}$20$}}%
      \put(3998,320){\makebox(0,0){\strut{}$24$}}%
      \put(892,2854){\makebox(0,0){\strut{}(a)}}%
    }%
    \gplgaddtomacro\gplfronttext{%
      \csname LTb\endcsname%
      \put(80,1794){\rotatebox{-270}{\makebox(0,0){\strut{}local density}}}%
      \put(2374,70){\makebox(0,0){\strut{}$x\, [\lambda/2]$}}%
      \csname LTb\endcsname%
      \put(2577,1338){\makebox(0,0)[r]{\strut{}without tuning}}%
      \csname LTb\endcsname%
      \put(2577,1159){\makebox(0,0)[r]{\strut{}tune $\omega$}}%
      \csname LTb\endcsname%
      \put(2577,980){\makebox(0,0)[r]{\strut{}tune $g$}}%
      \csname LTb\endcsname%
      \put(2577,801){\makebox(0,0)[r]{\strut{}tune both $g$, $\omega$}}%
      \csname LTb\endcsname%
      \put(2577,622){\makebox(0,0)[r]{\strut{}target state}}%
    }%
    \gplbacktext
    \put(0,0){\includegraphics{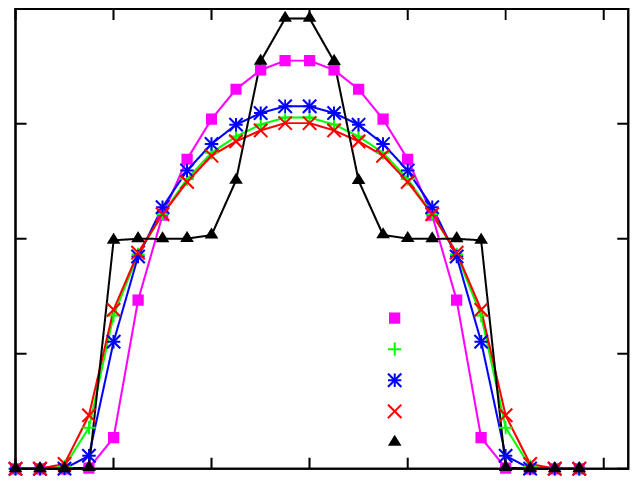}}%
    \gplfronttext
  \end{picture}%
\endgroup

\vspace{3mm}
% GNUPLOT: LaTeX picture with Postscript
\begingroup
  \makeatletter
  \providecommand\color[2][]{%
    \GenericError{(gnuplot) \space\space\space\@spaces}{%
      Package color not loaded in conjunction with
      terminal option `colourtext'%
    }{See the gnuplot documentation for explanation.%
    }{Either use 'blacktext' in gnuplot or load the package
      color.sty in LaTeX.}%
    \renewcommand\color[2][]{}%
  }%
  \providecommand\includegraphics[2][]{%
    \GenericError{(gnuplot) \space\space\space\@spaces}{%
      Package graphicx or graphics not loaded%
    }{See the gnuplot documentation for explanation.%
    }{The gnuplot epslatex terminal needs graphicx.sty or graphics.sty.}%
    \renewcommand\includegraphics[2][]{}%
  }%
  \providecommand\rotatebox[2]{#2}%
  \@ifundefined{ifGPcolor}{%
    \newif\ifGPcolor
    \GPcolortrue
  }{}%
  \@ifundefined{ifGPblacktext}{%
    \newif\ifGPblacktext
    \GPblacktexttrue
  }{}%
  % define a \g@addto@macro without @ in the name:
  \let\gplgaddtomacro\g@addto@macro
  % define empty templates for all commands taking text:
  \gdef\gplbacktext{}%
  \gdef\gplfronttext{}%
  \makeatother
  \ifGPblacktext
    % no textcolor at all
    \def\colorrgb#1{}%
    \def\colorgray#1{}%
  \else
    % gray or color?
    \ifGPcolor
      \def\colorrgb#1{\color[rgb]{#1}}%
      \def\colorgray#1{\color[gray]{#1}}%
      \expandafter\def\csname LTw\endcsname{\color{white}}%
      \expandafter\def\csname LTb\endcsname{\color{black}}%
      \expandafter\def\csname LTa\endcsname{\color{black}}%
      \expandafter\def\csname LT0\endcsname{\color[rgb]{1,0,0}}%
      \expandafter\def\csname LT1\endcsname{\color[rgb]{0,1,0}}%
      \expandafter\def\csname LT2\endcsname{\color[rgb]{0,0,1}}%
      \expandafter\def\csname LT3\endcsname{\color[rgb]{1,0,1}}%
      \expandafter\def\csname LT4\endcsname{\color[rgb]{0,1,1}}%
      \expandafter\def\csname LT5\endcsname{\color[rgb]{1,1,0}}%
      \expandafter\def\csname LT6\endcsname{\color[rgb]{0,0,0}}%
      \expandafter\def\csname LT7\endcsname{\color[rgb]{1,0.3,0}}%
      \expandafter\def\csname LT8\endcsname{\color[rgb]{0.5,0.5,0.5}}%
    \else
      % gray
      \def\colorrgb#1{\color{black}}%
      \def\colorgray#1{\color[gray]{#1}}%
      \expandafter\def\csname LTw\endcsname{\color{white}}%
      \expandafter\def\csname LTb\endcsname{\color{black}}%
      \expandafter\def\csname LTa\endcsname{\color{black}}%
      \expandafter\def\csname LT0\endcsname{\color{black}}%
      \expandafter\def\csname LT1\endcsname{\color{black}}%
      \expandafter\def\csname LT2\endcsname{\color{black}}%
      \expandafter\def\csname LT3\endcsname{\color{black}}%
      \expandafter\def\csname LT4\endcsname{\color{black}}%
      \expandafter\def\csname LT5\endcsname{\color{black}}%
      \expandafter\def\csname LT6\endcsname{\color{black}}%
      \expandafter\def\csname LT7\endcsname{\color{black}}%
      \expandafter\def\csname LT8\endcsname{\color{black}}%
    \fi
  \fi
    \setlength{\unitlength}{0.0500bp}%
    \ifx\gptboxheight\undefined%
      \newlength{\gptboxheight}%
      \newlength{\gptboxwidth}%
      \newsavebox{\gptboxtext}%
    \fi%
    \setlength{\fboxrule}{0.5pt}%
    \setlength{\fboxsep}{1pt}%
\begin{picture}(4320.00,3240.00)%
    \gplgaddtomacro\gplbacktext{%
      \csname LTb\endcsname%
      \put(550,470){\makebox(0,0)[r]{\strut{}$0$}}%
      \put(550,1000){\makebox(0,0)[r]{\strut{}$0.2$}}%
      \put(550,1530){\makebox(0,0)[r]{\strut{}$0.4$}}%
      \put(550,2059){\makebox(0,0)[r]{\strut{}$0.6$}}%
      \put(550,2589){\makebox(0,0)[r]{\strut{}$0.8$}}%
      \put(550,3119){\makebox(0,0)[r]{\strut{}$1$}}%
      \put(610,320){\makebox(0,0){\strut{}$0$}}%
      \put(1198,320){\makebox(0,0){\strut{}$50$}}%
      \put(1786,320){\makebox(0,0){\strut{}$100$}}%
      \put(2375,320){\makebox(0,0){\strut{}$150$}}%
      \put(2963,320){\makebox(0,0){\strut{}$200$}}%
      \put(3551,320){\makebox(0,0){\strut{}$250$}}%
      \put(4139,320){\makebox(0,0){\strut{}$300$}}%
      \put(904,2324){\makebox(0,0){\strut{}(b)}}%
    }%
    \gplgaddtomacro\gplfronttext{%
      \csname LTb\endcsname%
      \put(80,1794){\rotatebox{-270}{\makebox(0,0){\strut{}$|\langle\psi_{\rm target}|\psi_{\rm final}\rangle|$}}}%
      \put(2374,70){\makebox(0,0){\strut{}$t_R\, [\hbar / E_r]$}}%
    }%
    \gplbacktext
    \put(0,0){\includegraphics{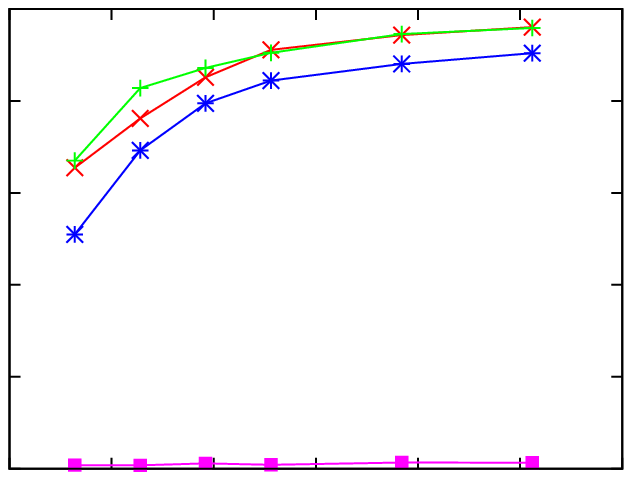}}%
    \gplfronttext
  \end{picture}%
\endgroup

\caption{Mott insulator with a metallic core: (a) Local density distribution profiles for the optimal initial states obtained from all lattice loading protocols. The target state is shown for reference. (b) Dependence of the fidelity on ramp time (starting from the optimal state) for all lattice loading methods.}
\label{fig:compare}
\end {figure}

We survey several values of the parameters $(g_i,\omega_i)$ in order to find a state that fits our two essential qualitative properties and thus turns out to be a good initial state. One candidate for an optimal initial state is obtained for $\omega_i = 0.16\, (\hbar/E_r)^{-1}, g_i = 1.4\, E_r \lambda/2$. In Fig. \ref{fig:compare}(a) we show the density profile for this optimal state along with the optimal states obtained from the other protocols. We compare the fidelity dependence on ramp time for the different protocols suggested in Fig. \ref{fig:compare}(b). We find that modified loading protocols achieve significant improvement in fidelity over the simple ramp of the lattice potential. Moreover all the three protocols are more or less equally efficient in doing so. This highlights once more that density defects are a relevant cause of heating that can be overcome by adjusting the model parameters to minimize variations in local density distribution with respect to the target state.

%%%%%%%%%%%%%%%%%%%%%%%%%%%%%%%%%%%%%%%%%%%%%%%%%%%%%%%%%%%%%%%%%%%%%%%%%%%%%%%%%%%%%%%%%%%%%%%%%%%%%%%%%

\section{Conclusions and Outlook}\label{sec:conclusions}
Our key result is that, similar to the bosonic case, density redistribution is  the main source of heating during optical lattice loading also for fermions. This indicates that modifying the loading scheme to keep the density distribution during the lattice loading similar to that of the desired target state can significantly reduce heating.

This can be achieved in various ways, for example by modifying the trapping or interaction strength during optical lattice loading.  Our numerical simulations show that these approaches are equally efficient, thus leaving room to select the one that is best suited to experimental setup. The fidelity of the final state is significant improved up to by a factor $\times 50$.

Although our numerical results are for one dimensional fermionic optical lattices, the conclusions carry over to higher dimensional systems. While DMRG methods are inefficient in higher dimensions, density profiles can be calculated using other numerical techniques, such as quantum Monte Carlo approaches. Adjusting system parameters to achieve similar density profiles throughout lattice loading can pave the way to lower temperatures in optical lattices.

%%%%%%%%%%%%%%%%%%%%%%%%%%%%%%%%%%%%%%%%%%%%%%%%%%%%%%%%%%%%%%%%%%%%%%%%%%%%%%%%%%%%%%%%%%%%%%%%%%%%%%%%%

% Acknowledgment
\begin{acknowledgments}
The simulations were performed using the ALPS MPS code~\cite{dolfi2014,bauer2011-alps,albuquerque2007} on the  M\"onch cluster of ETH Zurich.
This project was supported by the Swiss National Science Foundation through the National Center of Competence in Research Quantum Science and Technology QSIT and by ERC Advanced Grant SIMCOFE. MT acknowledges hospitality of the Aspen Center for Physics, supported by NSF grant PHY-1066293.
\end{acknowledgments}

\bibliography{bibliography}{}
\bibliographystyle{apsrev4-1}

\end{document}